\documentclass[twocolumn,pra,aps,superscriptaddress]{revtex4-2}

\usepackage[latin9]{inputenc}
\usepackage{mathtools}
\usepackage{amsbsy}
\usepackage{amssymb}
\usepackage{amstext}
\usepackage{bm}
\usepackage{xcolor}
\definecolor{darkblue}{rgb}{0, 0, 1}
\newcommand{\RN}[1]{%
	\textup{\uppercase\expandafter{\romannumeral#1}}%
}
\usepackage[unicode=true,pdfusetitle,
bookmarks=false,
breaklinks=false,pdfborder={0 0 1},backref=false,colorlinks=true,allcolors=darkblue]
{hyperref}
\hypersetup{
	bookmarksnumbered=false,bookmarksopen=false}
\makeatletter

\@ifundefined{textcolor}{}
{%
	\definecolor{BLACK}{gray}{0}
	\definecolor{WHITE}{gray}{1}
	\definecolor{RED}{rgb}{1,0,0}
	\definecolor{GREEN}{rgb}{0,1,0}
	\definecolor{BLUE}{rgb}{0,0,1}
	\definecolor{CYAN}{cmyk}{1,0,0,0}
	\definecolor{MAGENTA}{cmyk}{0,1,0,0}
	\definecolor{YELLOW}{cmyk}{0,0,1,0}
}

\newcommand{\beq}{\begin{equation}}
\newcommand{\eeq}{\end{equation}}
\newcommand{\beqa}{\begin{eqnarray}}
\newcommand{\eeqa}{\end{eqnarray}}

\usepackage{amsmath}
\usepackage{graphicx}
\usepackage{amssymb}
\usepackage{txfonts,color}
\makeatother

\begin{document}
\title{Quantum Active Learning}

\author{Yongcheng Ding}
\email{jonzen.ding@gmail.com}
	\affiliation{Institute for Quantum Science and Technology, Department of Physics, Shanghai University, Shanghai 200444, China}
\affiliation{Department of Physical Chemistry, University of the Basque Country UPV/EHU, Apartado 644, 48080 Bilbao, Spain}

\author{Yue Ban}
\affiliation{Departamento de F\'isica, Universidad Carlos III de Madrid, Avda. de la Universidad 30, 28911 Legan\'es, Spain}

\author{Mikel Sanz}

\affiliation{Department of Physical Chemistry, University of the Basque Country UPV/EHU, Apartado 644, 48080 Bilbao, Spain}
\affiliation{EHU Quantum Center, University of the Basque Country UPV/EHU, 48940 Leioa, Spain}
\affiliation{Basque Center for Applied Mathematics (BCAM), Alameda de Mazarredo 14, Bilbao, 48009, Spain}
\affiliation{IKERBASQUE, Basque Foundation for Science, Plaza Euskadi 5, Bilbao, 48009, Spain}

\author{Jos\'e D. Mart\'in-Guerrero}
\email{jose.d.martin@uv.es}
\affiliation{IDAL, Electronic Engineering Department,  ETSE-UV, University of Valencia, Avgda. Universitat s/n, 46100 Burjassot, Valencia, Spain}
\affiliation{Valencian Graduate School and Research Network of Artificial Intelligence (ValgrAI), Valencia, Spain}

\author{Xi Chen}
\email{xi.chen@ehu.eus}
\affiliation{Department of Physical Chemistry, University of the Basque Country UPV/EHU, Apartado 644, 48080 Bilbao, Spain}
\affiliation{EHU Quantum Center, University of the Basque Country UPV/EHU, 48940 Leioa, Spain}

\date{\today}

\begin{abstract}	
Quantum machine learning (QML), as an extension of classical machine learning that harnesses quantum mechanics, facilitates effiient learning from data encoded in quantum states. Training a quantum neural network (QNN) typically demands a substantial labeled training set for supervised learning. Human annotators, often experts, provide labels for samples through additional experiments, adding to the training cost. To mitigate this expense, there is a quest for methods that maintain model performance over fully labeled datasets while requiring fewer labeled samples in practice, thereby extending few-shot learning to the quantum realm. Quantum active learning (QAL) estimates the uncertainty of quantum data to select the most informative samples from a pool for labeling. Consequently, a QML model is supposed to accumulate maximal knowledge as the training set comprises labeled samples selected via sampling strategies. Notably, the QML models trained within the QAL framework are not restricted to specific types, enabling performance enhancement from the model architecture's perspective towards few-shot learning. Recognizing symmetry as a fundamental concept in physics ubiquitous across various domains, we leverage the symmetry inherent in quantum states induced by the embedding of classical data for model design. We employ an equivariant QNN capable of generalizing from fewer data with geometric priors. We benchmark the performance of QAL on two classification problems, observing both positive and negative results. QAL effectively trains the model, achieving performance comparable to that on fully labeled datasets by labeling less than 7\% of the samples in the pool with unbiased sampling behavior. Furthermore, we elucidate the negative result of QAL being overtaken by random sampling baseline through miscellaneous numerical experiments. Our work lays the groundwork for real-world applications of QML by addressing realistic experimental costs.
\end{abstract}

\maketitle

\section{Introduction}
Machine learning carries its own set of costs, often tied to the consumption of  computational resources and energy. Yet, the labeling process, a crucial aspect of training machine learning models, is frequently overlooked. This process demands the involvement of numerous human annotators, incurring significant financial and time inverstments. A fundamental question arises: Can machine learning models be effectively trained with only a few labeled samples, achieving comparable performance to models trained on orders of magnitude more samples? Few-shot learning emerges as a solution to this challenge~\cite{wang2020generalizing,song2023comprehensive}, encompassing various algorithmic~\cite{ravi2017optimization,du2020few,shen2021partial} and model-based approaches~\cite{,snell2017prototypical,garcia2017few,sung2018learning,zhang2018metagan}. Among these, active learning (AL) stands out as one of the most promising methodologies~\cite{settles2009active}. AL enhances model performance by iteratively labeling the most informative samples from a pool of unlabeled data, guided by the model's estimation. Recently, this framework and its application have been extended from computer science to physics~\cite{ding2023active}. This expansion is driven by the vast amount of experimental data generated in both laboratory settings and numerical simulations. AL can facilitate more efficient resource utilization and knowledge acquisition. Consequently, it has been applied in various subfields such as condensed matter physics~\cite{zhang2019active,yao2020active,tian2021efficient,sahinovic2021active,ding2022active,lysogorskiy2023active}, high-energy physics~\cite{caron2019constraining,diaw2020multiscale,rocamonde2022picking,mroczek2023mapping}, and quantum information~\cite{ding2020retrieving,zhu2022active,lange2023adaptive}, with the aim of reducing experiment costs.

After witnessing the success of AL in physics, we embark on exploring its application in quantum machine learning (QML)~\cite{biamonte2017quantum,cerezo2021variational}, by introducing the novel terminology  \textit{quantum active learning} (QAL) in the title. QML utilizes the unique properties of quantum systems to process quantum data, which may encode classical information, with remarkable efficiency, presenting practical advantages over classical methods~\cite{huang2022provably} in the noisy intermediate-scale era. Besides using quantum computing to optimize AL algorithms, the primary focus of QAL is to replace the classical machine learning model with a quantum neural network (QNN) in the workflow. Through QAL, it becomes feasible to refine the training of QML models by mapping their outputs to probabilistic distributions across all classes. To preliminarily verify feasibility,  recent experimental work on a photonic quantum processor has been carried out to train a simple quantum classifier with QAL~\cite{ding2023qst}. Indeed, QAL is not confined to any specific model,  offering the potential for enhancements to model performance through architectural designs aimed at facilitating few-shot learning.

In classical machine learning, one strategy to model architecture involves incorporating symmetry  through geometric deep learning~\cite{bronstein2021geometric}. Extending this concept to the design of equivariant QNNs have given rise to the prominent field of geometric quantum machine learning (GQML)~\cite{larocca2022group,meyer2023exploiting,schatzki2024theoretical,zheng2023speeding,nguyen2022theory,ragone2022representation,tuysuz2024symmetry}. Analogous to its classical counterpart, geometric prior is introduced as an inductive bias to enhance model performance, capitalizing on the pervasive presence of symmetry in quantum mechanics. In contrast to QNNs employing uniform inductive biases, GQML addresses barren plateaus~\cite{mcclean2018barren,marrero2021entanglement,holmes2022connecting,sack2022avoiding} by reducing expressibility, resulting in a more focused search space. Therefore, integrating GQML into the framework of QAL promises increased efficiency, facilitating advancements in few-shot learning within the quantum realm at both algorithmic and architectural levels.

Our primary contribution in this work lies in introducing QAL within the gate-model quantum computing paradigm, alongside the development of equivariant deep QNNs. To accommodate readers with limited computer science backgrounds, in Sec. \ref{preliminaries}, we first provide an explaination of  AL in the context of classical machine learning, followed by a definition of QNNs. We then delve into concise explanations of symmetry and equivariant QNNs, essential for grasping the concept of GQML in Sec. \ref{preliminaries}. With this foundational knowledge in place,  in Sec. \ref{slicing}, we illustrate QAL through the example of \textit{slicing-a-donut} with $\mathcal{Z}_2$ symmetry, presenting all aspects within a two-dimensional parameter space. Numerical experiments corroborate the positive impact of QAL and GQML on few-shot learning. In Sec. \ref{tic-tac-toe}, we tackle a more intricate problem involving $\mathcal{D}_4$ symmetry, akin to \textit{tic-tac-toe}, where we observe an understandable negative outcome.  We analyze this result, identifying the associated limitations of standard sampling strategies, and validate our findings through diverse numerical experiments. Finally, in Sec. \ref{Conclusion}, we offer insights into future research direction aimed at optimizing and customizing QAL for diverse quantum data and tasks by integrating complementary methods. We believe that QAL holds promise as a framework for reducing the overall cost of analyzing quantum experiment results and designing quantum experiments via QML.

\section{Theoretical framework}
\label{preliminaries}

We hereby provide the definitions and preliminary knowledge required for understanding our framework.

\subsection{Active learning}
AL operated under the hypothesis that a machine learning model can be efficiently trained using only a small subset of samples from a larger pool of unlabeled data. These selected samples should be highly representative and informative, contributing significant knowledge when labeled by annotators.
 
To formulate  problem broadly, we define a pool $\mathcal{U}=\{\textbf{x}_i\}_{i=1}^N$  consisting of $N$ unlabeled samples $\textbf{x}_i$, each represented as a multi-dimensional vector from the domain $\mathcal{X}$.  The true labels $y_i$ from the domain $\mathcal{Y}$ of $\textbf{x}_i$ are unknown, which can be obtained by annotators after experiments. Meanwhile, we have another set of samples $\mathcal{T}=\{(\textbf{x}_j,y_j)\}_{j=1}^M$ serving as the test set. Both datasets, $\mathcal{U}$ with true labels and $\mathcal{T}$, follow a distribution over $\mathcal{X}\times\mathcal{Y}$. We assume the existence of  an unknown function $f:\mathcal{X}\rightarrow\mathcal{Y}$ that maps each sample to its label, i.e., $f(\textbf{x}_i)=y_i$. The goal is to train a parameterized model $h_\theta$ that approximates $f$ well using a significantly smaller training set $\mathcal{\tilde{U}}=\{(\textbf{x}_i,y_i)\}_{i=1}^{\tilde{N}}$, $\tilde{N}\ll N$, selected by $h_\theta$ itself. The quality of approximation is evaluated by comparing the model's prediction on the testing set $h_\theta(\textbf{x}_j)$ with the corresponding true labels $y_j$.

The most common strategy, known as uncertainty sampling (USAMP)~\cite{lewis1995sequential}, requires mapping the model $h_\theta$ to a probabilistic model $P_\theta$ that evaluates the probability $P_\theta(y_j|\mathbf{x})$ of $\mathbf{x}$ belonging to the $j$-th class. For binary classification problem, it queries the sample whose probability is the closest to fifty-fifty. For multinomial classification, we introduce the following definition.

\textbf{Definition 1:} (Least confidence). The model queries the unlabeled sample $\textbf{x}_{\text{LC}}$  that satisfies:
\begin{eqnarray}
\textbf{x}_{\text{LC}} &=& \underset{\textbf{x}}{\text{argmax}}\left[1-P_\theta(\hat{y}|\textbf{x})\right],\nonumber\\
\hat{y} &=& \underset{y}{\text{argmax}}\left[P_\theta(y|\textbf{x})\right].
\end{eqnarray}
Here, $\hat{y}$ is the most probable label of $\textbf{x}$ estimated by $h_\theta$, which is also the only class information considered by the strategy for sampling. To optimize the strategy for multinomial classification, researchers propose a variant~\cite{scheffer2001active} defined below.

\textbf{Definition 2:} (Margin sampling). The model queries the unlabeled sample $\textbf{x}_{\text{MS}}$ that fulfills 
\begin{equation}
\textbf{x}_{\text{MS}} = \underset{\textbf{x}}{\text{argmin}}\left[P_\theta(y_1|\textbf{x})-P_\theta(y_2|\textbf{x})\right],
\end{equation}
where $y_1$ and $y_2$ are the first and second most likely labels of \textbf{x}.

The insight is that samples with small margins exhibit greater ambiguity.
Accurately discerning the true label in such cases becomes crucial for enhancing the model's discriminatory capability between them. Consequently, it is natural to consider the probability of all classes in the formulation, aligning with Shannon's  theory~\cite{shannon1948mathematical}.

\textbf{Definition 3:} (Entropy sampling). The model queries the unlabeled sample $\textbf{x}_{\text{ES}}$ that satisfies
\begin{equation}
\textbf{x}_{\text{ES}} = \underset{\textbf{x}}{\text{argmax}}\left[-\sum_jP_\theta(y_j|\textbf{x})\log P_\theta(y_j|\textbf{x})\right].
\end{equation}
Entropy serves as an information-theoretic gauge, quantifying the information required to represent a distribution. It is commonly interpreted as a metric reflecting uncertainty or impurity in machine learning. Thus, it inspires another query selection framework called Query-By-Committee (QBC)~\cite{settles2008active}, which considers information from multiple models instead of multiple classes. In this framework, a committee $\mathcal{C}= {h_\theta^{(1)}, \ldots, h_\theta^{(C)}}$ that consists of $C$ competing hypotheses is proposed to approximate the function $f$ individually. Thus, there is also an entropy-like definition for sampling as follows.

\textbf{Definition 4:} (Voting entropy). The committee of competing models queries the unlabeled sample $\textbf{x}_{\text{VE}}$ that satisfies
\begin{equation}
\textbf{x}_{\text{VE}} = \underset{\textbf{x}}{\text{argmax}}\left[-\sum_j \frac{V(y_j)}{C}\log\frac{V(y_j)}{C}\right],
\end{equation}
where $V(y_j)$ is the number of votes that a class receives from the committee members.

As observed, all these sampling strategies are \textit{model-based}, meaning they only focus on the estimation made by the model. In certain scenarios, this  problem can be problematic particularly when outlier samples with the least certainty reside on the classification boundary.  However, this approach may not accurately reflect the characteristics of other samples in the distribution, making it improbable that knowing its label would enhance the overall accuracy on the testing set. This issue can be addressed by explicitly modeling the distribution in the strategy design in terms of similarity. The central concept is that informative samples should not just be uncertain, they should also be indicative of the underlying distribution, meaning they are located in dense regions of the input space $\mathcal{X}$. Thus, a \textit{data-based} strategy can be defined.

\textbf{Definition 5:} (Density-weighted sampling). The model queries the unlabeled sample $\textbf{x}_{\text{DS}}$ that satisfies
\begin{equation}
\textbf{x}_{\text{DS}} = \underset{\textbf{x}}{\text{argmax}}\left[\frac{1}{U}\sum\text{Sim}(\textbf{x},\textbf{x}_{(u)})\right],
\end{equation} 
where $U<N$ is the size of the unlabeled pool, and $\textbf{x}_{(u)}$ are samples in the pool except $\textbf{x}$ itself. The model-based and data-based sampling can be combined to query samples for a better performance.

\subsection{Quantum neural network}

QNN, a parameterized unitary applicable in quantum devices, serves as the cornerstone of QAL. For classical data, it consists of unitary ansatze $U(\theta)$ with trainable parameters $\theta$ and encoders $E(\textbf{x})$ for data loading, followed by measurements of observables $\hat{\mathcal{O}}$. To overcome the no-cloning theorem limitation, data re-uploading, a technique introduced in~\cite{perez2020data,schuld2021effect}, effectively enhances the expressivity of QNN by encoding data into the quantum circuit multiple times.  The prediction on $\textbf{x}$ is calculated as
\begin{eqnarray}
h_\theta(\textbf{x}) &=& \text{Tr}[U_d(\theta)E(\textbf{x})\cdots U_1(\theta)E(\textbf{x})|0\rangle\langle0|^{\otimes k}\nonumber\\
&~&E^\dag(\textbf{x})U_1^\dag(\theta)\cdots E^\dag(\textbf{x})U_d^\dag(\theta)\hat{\mathcal{O}}],
\end{eqnarray}
where $d$ represents the number of encoder-ansatz layers. Thus, the loss function $\mathcal{L}(h_\theta)$, constructed with the expectation on observables $\langle\hat{\mathcal{O}}\rangle$, is minimized by classical optimizers, thereby fitting the function for mapping the distribution $f:\mathcal{X}\rightarrow\mathcal{Y}$.

Furthermore, QNN can be employed to analyze quantum data $\rho$ represented as a density matrix from quantum experiments. In this scenario, it takes the form
\begin{eqnarray}
h_\theta(\rho)&=&\text{Tr}[U(\theta)\rho^{\otimes k}U^\dag(\theta)\hat{\mathcal{O}}],\nonumber\\
&=&\text{Tr}[\mathcal{W}_\theta(\rho^{\otimes k})\hat{\mathcal{O}}]
\end{eqnarray}
where $k$ denotes the number of copies of quantum data $\rho$, and $\mathcal{W}_\theta$ represents a trainable quantum channel with parameters $\theta$. For $k=1$, each $\rho_i$ is processed and measured individually. Conversely, for $k\geq 2$, results of multiple but same quantum experiments are stored in quantum memory, being processed and measured concurrently. Notably, in QNN for quantum data, the data re-uploading layers are not present, while ansatz layers can be expressed by one unitary operator. Nonetheless, it is interesting to point out that QNNs for classical and quantum data share the same framework if the data is loaded only once, i.e., $E(\textbf{x})$ serves as the quantum state preparation, mapping $\textbf{x}\in\mathcal{X}$ to $\rho$ in the Hilbert space of the same dimension.

\begin{figure}
\includegraphics[width=0.95\linewidth]{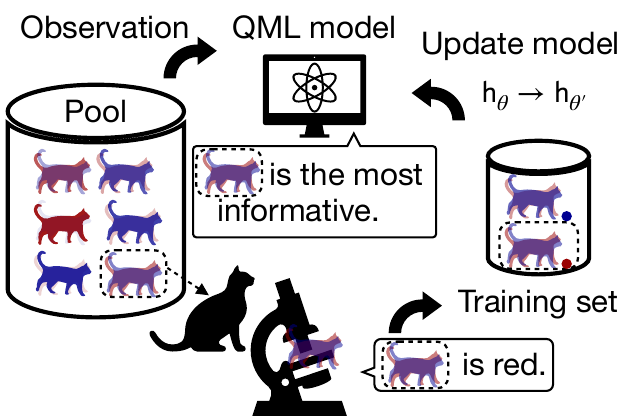}
\caption{\label{fig:scheme_qal} Schematic diagram of QAL. Quantum data (or classical data encoded in quantum states) are observed (or uploaded) by the QML model $h_\theta$. The model estimates which sample will introduce more knowledge after being labeled by the expert annotator with experiments. The annotator labels the queried sample and sends it to the training set for supervised learning, updating the parameters of the QML model.
}
\end{figure}

Combining the principles of AL with QNN opens the door to QAL, a promising framework (illustrated in Fig.~\ref{fig:scheme_qal}) for enhancing model efficiency with limited quantum and classical labeled data. Thanks to the nature of quantum mechanics, it is trivial to map a QNN to a probabilistic model. For example, for $\langle\hat{Z}\rangle$, the squared amplitudes on $|0\rangle$ and $|1\rangle$ are indeed the probabilities of the label estimation.

\subsection{Geometric quantum machine learning}
\begin{figure}
\includegraphics[width=0.95\linewidth]{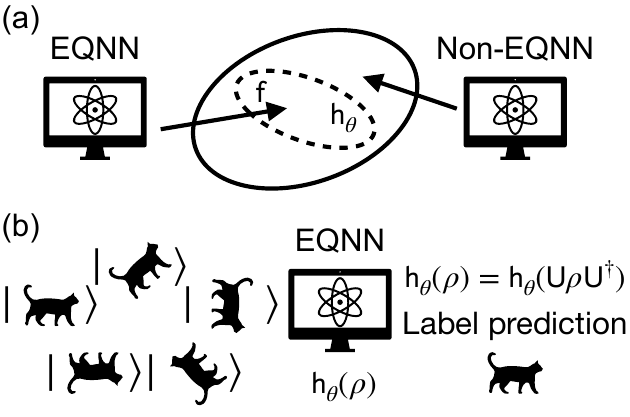}
\caption{\label{fig:scheme_gqml} (a) Equivariant QNN $h_\theta$ with geometric priors reduces the search space for approximating the label mapping function $f$ by exploiting the symmetry of the quantum data. (b) To design equivariant QNNs, the quantum data (or classical data after encoding) preserve the properties that determine the label after transformations.}
\end{figure}

As depicted in Fig.~\ref{fig:scheme_gqml}(a), GQML is a scheme dedicated to designing QNNs with inductive biases based on the symmetry of the data to be learned. QNNs with geometric priors offer generalization from very few data, making them particularly advantageous when the number of labeled samples is limited.

The underlying philosophy of GQML is that the data, either in the classical realm or from quantum experiments, often exhibits symmetry, i.e., the properties remain after certain transformation [see Fig.~\ref{fig:scheme_gqml}(b)]. These properties can be inherently linked to the label of the data, giving rise to the following definitions.

\textbf{Definition 5:} (Label invariance in quantum data). $\mathcal{G}$ is a group such that $\forall U \in\mathcal{G}$, some properties of $U\rho U^\dag$ remain unchanged. The group introduces label invariance if
\begin{equation}
f(U\rho_iU^\dag)=f(\rho_i)=y_i,
\end{equation} 
for all $\rho_i$ that belong to label $y_i$.

\textbf{Definition 6:} (Label invariance in classical data). $\mathcal{S}$ is a symmetry group with representation $R$. Label invariance is introduced if
\begin{equation}
f(\rho_{R(g)\cdot\textbf{x}_i}) = f(\rho_{\textbf{x}_i}) = y_i,~\forall g\in\mathcal{S},
\end{equation}
where $\rho_{\textbf{x}_i}$ is the result of the encoding of classical data $\Psi(\textbf{x}_i):\textbf{x}_i\rightarrow\rho_{\textbf{x}_i}$.

For example, the winner of a chess-like game is not changed if the chessboard is rotated or flipped. In quantum mechanics, the purity of a quantum state in a $d$-dimensional Hilbert space is preserved under a unitary operation $U(d)$. Similarly, the degree of entanglement, such as entangling entropy, remains unchanged after a local operation or SWAP. In other words, if the label of a sample depends on these properties, we can construct a QML model $h_\theta$ that guarantees label invariance under any action of group elements.

In $k$-copy quantum experiments, equivariance conditions should be satisfied to ensure that $h_\theta$ is invariant under $\mathcal{G}$:
\begin{equation}
\mathcal{W}_\theta\left[U^{\otimes k}\rho_i^{\otimes k}{(U^\dag)}^{\otimes k}\right] = U^{\otimes k}\mathcal{W}_\theta\left[\rho_i^{\otimes k}\right]{(U^\dag)}^{\otimes k},~\forall U\in\mathcal{G},
\end{equation}
\begin{equation}
[\hat{\mathcal{O}},U^{\otimes k}] = 0,~\forall U\in\mathcal{G}.
\end{equation}
Such equivarance can be linked to the classical data with equivariant encoding $\Psi$ with respect to the group element $g$ if and only if there exists an encoding-induced unitary representation $R_q(g)$ satisfying
\begin{equation}
\label{eq:equi-encoding}
\rho_{R(g)\cdot\textbf{x}_i} = R_q(g)\rho_{\textbf{x}}R_q^\dag(g).
\end{equation}
Thus, in the classical realm, one can capture the encoding-induced symmetry by equivariant quantum gates. We assume that the quantum gates generated by a Hermitian generator $G$ satisfies
\begin{equation}
\label{eq:UG}
U_G(\theta) = \exp(-i\theta G),~G\in\mathcal{G},
\end{equation}
leading us to the equivariance condition as follows.

\textbf{Definition 7:} (Gate invariance). The gate $U_G(\theta)$ generated by $G$ is equivariant with respect to symmetry $\mathcal{S}$, iff
\begin{equation}
\label{def:gate}
[U_G(\theta),R_q(g)] = 0,~\forall\theta\in\mathbb{R},~\forall g\in\mathcal{S}.
\end{equation}
One of the most practical methods for constructing the equivariant gateset is the so-called Twirling formula, given in the following definition.

\textbf{Definition 8:} (Twirling formula). Let $\mathcal{S}$ be the symmetry and $R$ be a representation of $\mathcal{S}$.
\begin{equation}
\label{def:twirling}
T_R[G]=\frac{1}{\mathcal{S}}\sum_{g\in\mathcal{S}}R(g)GR^\dag(g),
\end{equation}
defines a twirl onto the set of operators $[T_R[G],R(g)]=0,~\forall X$ and $g\in\mathcal{S}$.

In this manner, we move from the gate $U_G(\theta)=\exp(-i\theta G)$ to the twirled $U'=\exp(-i\theta T_R[G])$, ensuring that a generator commutes with a given representation. While this subsection provides a brief introduction to the critical background of GQML, interested readers can explore additional details on invariances, equivariance, proofs, and other methodologies of this field in pedagogical literature~\cite{larocca2022group,meyer2023exploiting,ragone2022representation}.

\section{Slicing a donut: $\mathcal{Z}_2$ symmetry}
\label{slicing}

Starting from this section, we employ QAL with equivariant QNN as models to demonstrate optimal training strategies with limited labels. The standard workflow begins with problem formulation, aiming to approximate the mapping $f:\mathcal{X}\rightarrow\mathcal{Y}$ with a QNN $h_\theta$. Designing the network structure requires studying the label symmetry to derive the equivariant gate set and corresponding measurements. Once the equivariant QNN is constructed, the model can evaluate the uncertainty of all unlabeled samples in the training pool by mapping the output to a probabilistic distribution over all classes. The most informative sample is then queried using sampling strategies, labeled by annotators, and added to the training set to update the model parameters $\theta$. This iterative process allows training the QNN with a smaller training set, achieving satisfactory model performance at a lower cost by avoiding the need to label all samples in the training pool. To illustrate the proposal, we begin by considering a toy model, which is the game between Alice and Bob called \textit{slicing-a-donut}.

\subsection{Problem formulation}
\begin{figure}
\includegraphics[width=1\linewidth]{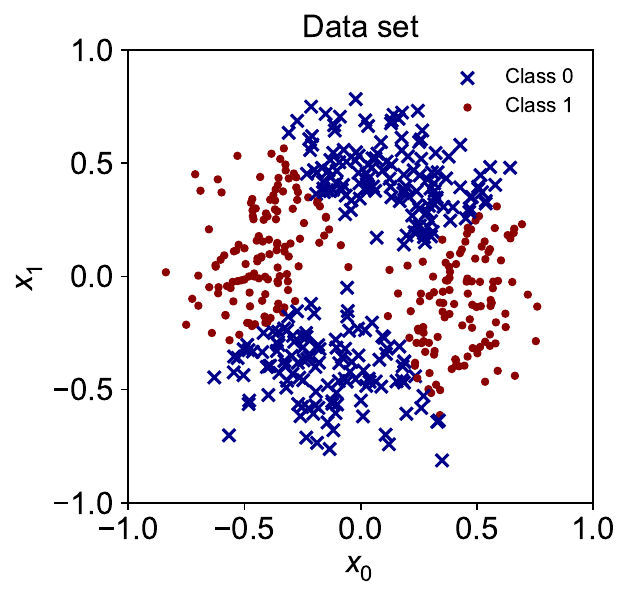}
\caption{\label{fig:dataset} Donut-like dataset with $\mathcal{Z}_2$ label symmetry. We generate 500 samples by mapping the polar coordinates $(r_i,\theta_i)$ to $\text{x}_i=(x_0,x_1)$, where the radius and polar angle follow Gaussian distributions $G(\mu=0.5,\sigma=0.15)$ and uniform distribution $[0,2\pi]$, respectively. To ensure $\mathcal{Z}_2$ label symmetry, we calculate $\cos(2\theta_i+0.58)$ and label the sample as class 0/1 if the value is negative/positive. Later, the dataset is randomly separated by a ratio of $3:1:1$ as a training pool, validation set, and test set for the game between Alice and Bob.}
\end{figure}

Alice creates an ad-hoc dataset of samples, where the spatial distribution in parameter space resembles a donut. The labels of the samples follow $\mathcal{Z}_2$ symmetry, i.e., $y_i=f(\textbf{x}_i)=f[R(\sigma)\cdot\textbf{x}_i]=f(-\textbf{x}_i)$, where $\sigma$ is a non-identical element of the $\mathcal{Z}_2$ group. The full dataset (see Fig.~\ref{fig:dataset}) is randomly separated into the training pool, validation set, and test set according to a ratio of $3:1:1$ for further processing. Alice invites Bob to train a QML model that slices the donut for binary classification, but with limited samples from the training pool for labeling. Bob knows the locations of the samples in the parameter space but has no information on the labels. However, Bob can query the labels of selected samples from Alice for supervised learning.

\subsection{QML model}
\begin{figure}
\includegraphics[width=0.95\linewidth]{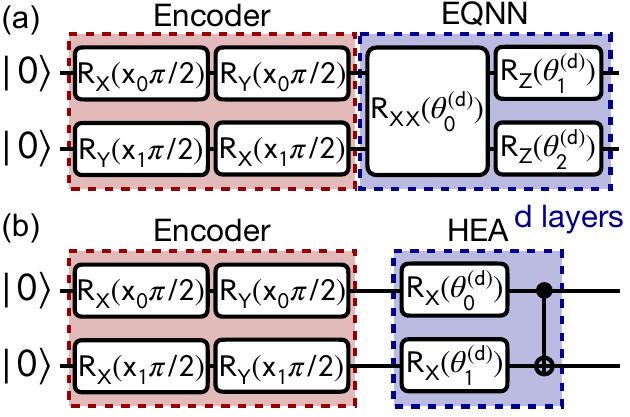}
\caption{\label{fig:qnnz2} (a) QNN structure of EQNN-Z designed with geometric priors for QML and (b) HEA designed with pseudo-symmetry as the baseline. Both QNNs load classical data only once with the encoder at the beginning. After repeating $d$ layers of the ansatz, the expectation of the equivariant operator $\langle\hat{\mathcal{O}}\rangle=(\hat{Z}_0+\hat{Z}_1)/2$ is measured as the output of the model.}
\end{figure}

Now we introduce the QML model Bob used in numerical experiments. To combine QAL with GQML, Bob chooses the state-of-the-art equivariant network called EQNN-Z, as shown in Fig.~\ref{fig:qnnz2}(a), which was initially proposed in preliminary work for a different dataset~\cite{tuysuz2024symmetry}. To encode the classical data into a quantum state, the model starts from the encoding layer with rotation gates $R_X$ and $R_Y$. We highlight that here, the orders of $R_X$ and $R_Y$ are different on the first and second qubits to break an undesired pseudo-symmetry at the encoding level. Specially, if $R_X$ and $R_Y$  are applied in the same order,  there exists the SWAP pseudo-symmetry $x_i^0\leftrightarrow x_i^1$ in the wave function after data encoding. This would result in the same expectation if we choose $\langle\hat{\mathcal{O}}\rangle=\langle\hat{Z}_0+\hat{Z}_1\rangle/2$ as the observable to be measured.

Once the data encoding is chosen, it spontaneously induces the unitary representation of group element $\sigma$. We denote the encoder $\Psi(\textbf{x}_i):\textbf{x}_i\rightarrow\rho_{\textbf{x}_i}$ by unitary operator $U(x_i^0,x_i^1)=[R_Y(x_i^0\cdot\pi/2)\otimes R_X(x_i^1\cdot\pi/2)][R_X(x_i^0\cdot\pi/2)\otimes R_Y(x_i^1\cdot\pi/2)]$. Accordingly, we notice the following equation
\begin{equation}
U(-x_i^0,-x_i^1) = (\hat{Z}\otimes\hat{Z})U(x_i^0,x_i^1)(\hat{Z}\otimes\hat{Z}),
\end{equation}
where $\hat{Z}\otimes\hat{Z}=R_q(\sigma)$ is the encoding-induced unitary representation of non-identical group element $\sigma$, satisfying the condition for equivariant encoding~\eqref{eq:equi-encoding}.

To construct the equivariant ansatz, we choose three generators $G\in{\hat{Z}\otimes I, I\otimes\hat{Z},\hat{X}\otimes\hat{X}}$ that commute with the unitary representation $R_q(\sigma)=\hat{Z}\otimes\hat{Z}$, since if $[G,R_q(g)]=0$, then all powers of $G$ commute with $R_q(g)$, satisfying \textbf{Definition 7}, see Eq. \eqref{def:gate} for gate invariance. One can also verify this using the twirling formula \eqref{def:twirling}, yielding 
\begin{eqnarray}
T_R[G]=\frac{1}{2}\left[I_{4\times4}GI_{4\times4}+(\hat{Z}\otimes\hat{Z})G(\hat{Z}\otimes\hat{Z})\right]=G,
\end{eqnarray}
where the twirled generator $T_R[G]$ is not changed. Thus, we have single-qubit and two-qubit gates, see Eq.~\eqref{eq:UG},
\begin{eqnarray}
R_Z(\theta)&=&\exp(-i\theta \hat{Z}),\nonumber\\
R_{XX}(\theta)&=&\exp[-i\theta(\hat{X}\otimes\hat{X})],
\end{eqnarray}
leading us to the $d$-time repeated ansatz layers with $3d$ independent parameters
\begin{equation}
U_{\text{EQNN}}(\boldsymbol{\theta}^{(d)}) = [R_Z(\theta_1^{(d)})\otimes R_Z(\theta_2^{(d)})]R_{XX}(\theta_0^{(d)}).
\end{equation}
With the equivariant operator to be measured, i.e., $(\hat{Z}\otimes\hat{Z})(\hat{Z}_0+\hat{Z}_1)(\hat{Z}\otimes\hat{Z})/2=(\hat{Z}_0+\hat{Z}_1)/2$, we ensure that EQNN-Z preserves label symmetry by leveraging the geometric prior in the network structure.

At the same time, we need to construct a model without geometric prior that maximizes expressivity by employing a hardware-efficient ansatz (HEA). We introduce the undesired pseudo-symmetry by ordering the $R_X$ and $R_Y$ gates in the same manner on two qubits. The HEA consists of two independent parameterized $R_X$ gates that do not belong to the equivariant gate set, and a CNOT gate for entangling two qubits.
\begin{equation}
U_{\text{HAE}}(\boldsymbol{\theta}^{(d)}) = \text{CNOT}[R_X(\theta_0^{(d)})\otimes R_X(\theta_1^{(d)})].
\end{equation}
The operator to be measured is the same as those in EQNN-Z. This QNN, see Fig.~\ref{fig:qnnz2}(b), should also be trained as a baseline to validate GQML and study the choice for models in QAL.

\subsection{Training of the model}

\begin{figure}
\includegraphics[width=0.95\linewidth]{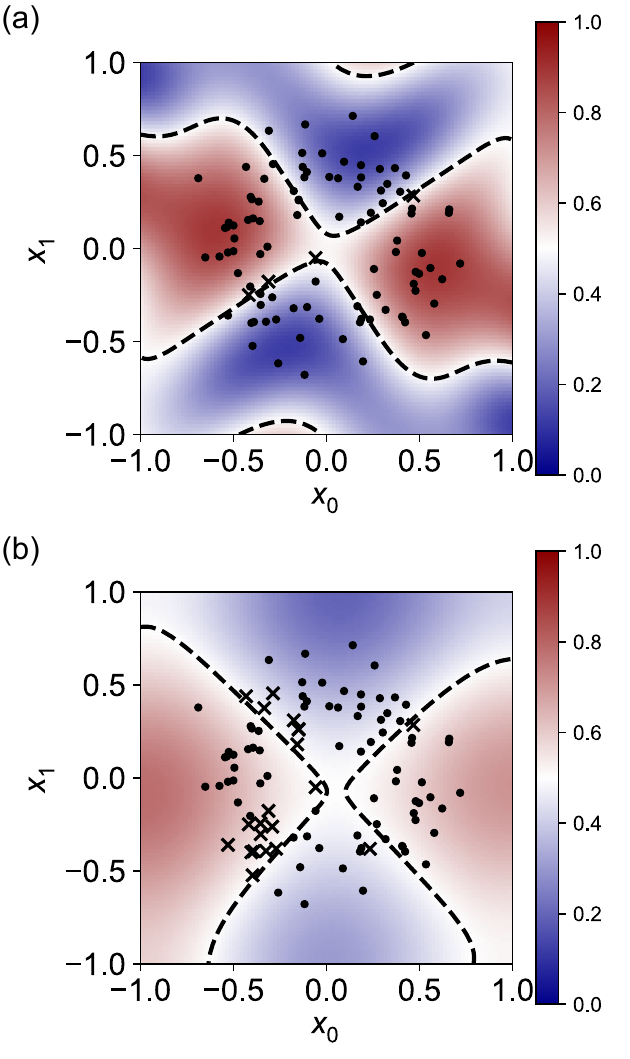}
\caption{\label{fig:supervised} Model estimation over parameter space and decision boundaries of (a) EQNN-Z with geometric priors and (b) HEA as the baseline QNN. These models are trained with full labels of all samples in the training pool. Correct and incorrectly classified samples in the test set are plotted as dots and crosses, respectively. Dashed curves characterize the decision boundaries for slicing the donut.}
\end{figure}

The assumption of QAL is that one can achieve satisfactory model performance by labeling the most informative samples estimated by QNNs for few-shot learning. Before implementing QAL, we need to verify if the model can classify two classes with full information, meaning that all samples in the training pool are labeled and used for model training. All parameters in QNNs are randomly initialized following a uniform distribution within $[-\pi,\pi]$. We use binary cross-entropy as the loss function
\begin{equation}
\mathcal{L}(\tilde{\mathcal{U}},\boldsymbol{\theta}) = -\frac{1}{|\tilde{\mathcal{U}}|}\sum_{y_i\in\tilde{\mathcal{U}}}\left[y_i\log\langle\hat{y_i}\rangle + (1-y_i)\log(1-\langle\hat{y_i}\rangle)\right],
\end{equation}
where $\tilde{\mathcal{U}}$ is the training set, $y_i = 0~\text{or}~1$ is the true label of sample $\textbf{x}_i$, and $\langle \hat{y_i}\rangle=(\langle\hat{\mathcal{O}}\rangle+1)/2$ is the renormalized output of QNN based on the expectation value of the measured operator $\hat{\mathcal{O}}=\langle\hat{Z}_0+\hat{Z}_1\rangle/2\in[-1,1]$. We optimize the loss function with the Adam optimizer using a learning rate of $0.1$ for $100$ epochs without batch method. While minimizing the loss function, we monitor the classification accuracy on the validation set and record the best accuracy in history to retrieve the model for evaluation on the test set. In Fig.~\ref{fig:supervised}(a), we display the model's estimation over the parameter space and the decision boundary between the two classes, achieving a satisfactory accuracy of $96\%$ with $d=3$ layers of equivariant ansatz, preserving the rigid $\mathcal{Z}_2$ label symmetry. For a fair comparison between EQNN-Z and HEA, we must limit the circuit depth to the same level. Note that $R_{XX}(\theta)$ as M\o lmer-S\o rensen gate is not directly implementable in all quantum platforms. Thus, we decompose it into $R_{XX}(\theta)=(H\otimes H)\text{CNOT}[I\otimes R_Z(2\theta)]\text{CNOT}(H\otimes H)$ for maximal compatibility. By only counting the number of two-qubit gates after decomposition for circuit depth, we set the layers of HEA to $d=6$ in numerical experiments for benchmarking, resulting in an example of $79\%$ accuracy in Fig.\ref{fig:supervised}(b). After training EQNN-Z and HEA with 40 different random initializations over full samples in the training pool, we achieve an average accuracy of $(91.63\pm4.24\%)$ and $(77.30\pm2.06\%)$, respectively, evaluated once the best historical performance on validation sets is observed after training for $49.88\pm28.62$ and $34.40\pm15.23$ epochs, proving that our hyperparameters are chosen adequately for model convergence.

\subsection{Numerical experiments}

\begin{figure}
\includegraphics[width=0.95\linewidth]{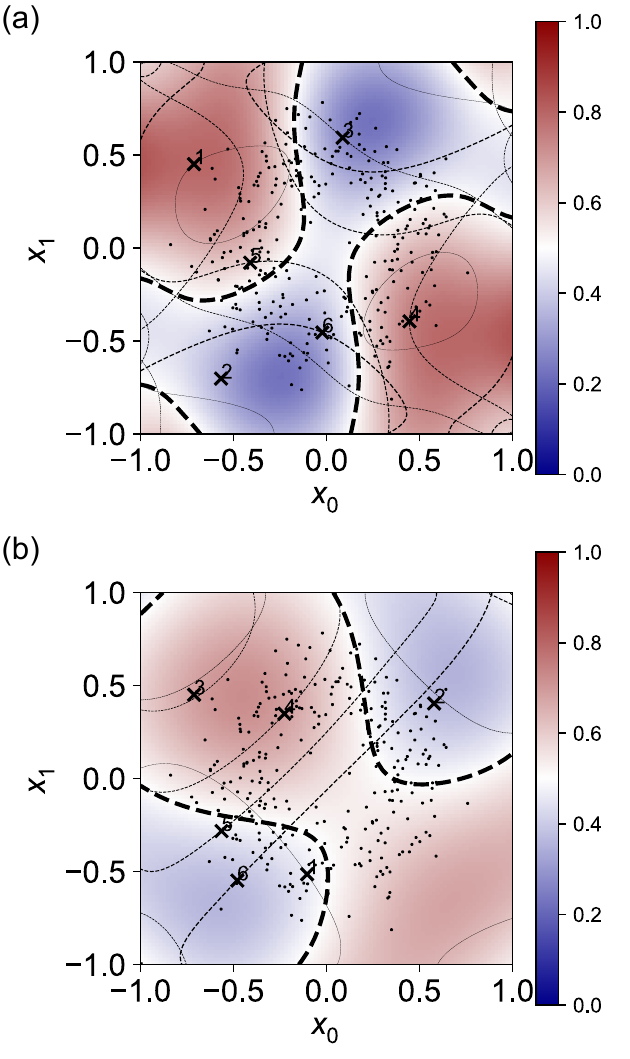}
\caption{\label{fig:sampling} Sampling behavior of USAMP with (a) EQNN-Z and (b) HEA. We plot the change in the decision boundary after each label query, from thin to thick lines. The location and order of queried samples are marked by crosses, while other available samples in the training pool are plotted as dots.}
\end{figure}

After demonstrating that GQML can achieve high accuracy with sufficient labeled samples, Bob initiates slicing-the-donut using QAL with a model-based-only USAMP strategy. Note that for binary classification, the three sampling strategies in \textbf{Definition 1-3} are equivalent; that is, the sample will be queried if it is nearest to the decision boundary among others. Based on the loss function, Bob defines the model's confidence as $|\langle\hat{\mathcal{O}}\rangle_{\textbf{x}}|$, and queries the sample $\textbf{x}_{QAL}={\text{argmax}_{\textbf{x}}}|\langle\hat{\mathcal{O}}\rangle_{\textbf{x}}|$ for its true label from Alice. In Fig.~\ref{fig:sampling}, we illustrate the sampling behavior of USAMP with EQNN-Z and HEA by recording the changes in the decision boundary and corresponding queried samples, achieving accuracies of $95\%$ and $69\%$ on the test set, respectively. Remarkably, in the case of EQNN-Z, Bob achieves the most satisfying accuracy with the combination of QAL and GQML in the game by only querying $6$ labels from Alice, approaching the theoretical maximum estimated by preliminary experiments.

\begin{figure}
	\includegraphics[width=0.95\linewidth]{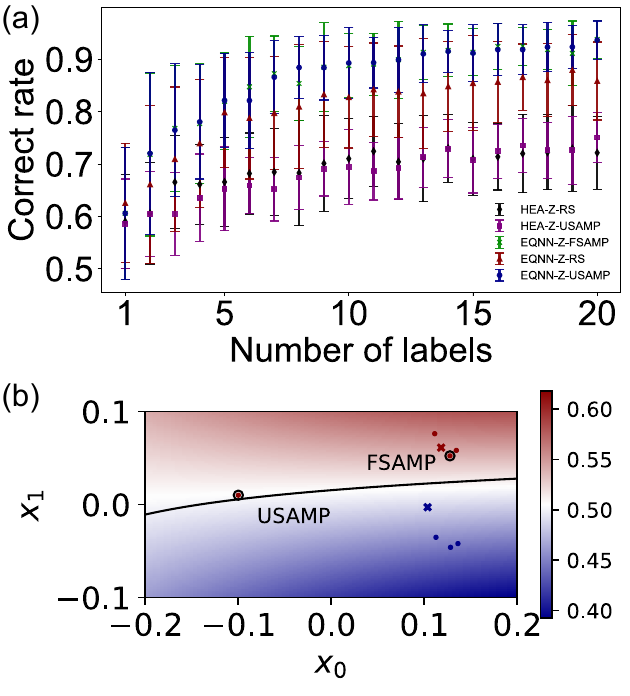}
	\caption{\label{fig:benchmark} (a) Mean correct rate of quantum machine learning models with random sampling, USAMP, and fidelity sampling as different sampling strategies for solving \textit{slicing-a-donut}. The error bars denote one standard deviation. (b) Corrected sampling behavior over an ad-hoc dataset. The model is $d=6$ HEA-Z trained with only two queried samples (red and blue crosses). According to USAMP, the circled sample on the left is closest to the decision boundary. However, querying the circled sample on the right according to FSAMP~\eqref{eq:FSAMP} (hyperparameter $\lambda=1$), by considering data distribution, is likely to introduce more information.}
\end{figure}

Once we fix the circuit depth and the number of feasible labels, we can fairly compare different sampling strategies and network structures. In Fig.~\ref{fig:benchmark}(a), we present a quantitative benchmark of our combination of QAL and GQML against random sampling and networks without geometric priors as baselines. The most convincing evidence of the validity of QAL is that USAMP outperforms random sampling with EQNN-Z, with the difference between accuracies meeting one standard deviation. We also extend \textbf{Definition 5} to the quantum version by defining the similarity as fidelity $\text{Sim}(\textbf{x},\textbf{x}_{(u)})=\left[\text{Tr}(\sqrt{\sqrt{\rho_{\textbf{x}}}\cdot\rho_{\textbf{x}_{(u)}}\cdot\sqrt{\rho_{\textbf{x}}}})\right]^2$ after encoding. Note that such data-based sampling will behave like random sampling in this specific problem since the samples in the training pool are dense and almost uniform in the angular direction. Thus, we have the fidelity sampling strategy as a combination of USAMP and density-weighted sampling (see green crosses)

\begin{equation}
\label{eq:FSAMP}
\resizebox{0.91\hsize}{!}{%
$\textbf{x}_{\text{FSAMP}} = \underset{\textbf{x}}{\text{argmin}}\left\{|\langle\hat{\mathcal{O}}\rangle_{\textbf{x}}-0.5| + \frac{\lambda}{U}\sum\left[1-\text{Tr}(\sqrt{\sqrt{\rho_{\textbf{x}}}\rho_{\textbf{x}_{(u)}}\sqrt{\rho_{\textbf{x}}}}\right]^2\right\},$
}
\end{equation}

where $\lambda=0.1$ is the hyperparameter in numerical experiments. One may notice that the performance of USAMP and fidelity sampling are hardly distinguishable since there is no exotic sample in the training set. In other words, fidelity sampling can only correct the sampling behavior that leads to better classification [see Fig.~\ref{fig:benchmark}(b)], a problem not encountered with our dataset. Meanwhile, the performance of USAMP and random sampling is also indistinguishable since the HEA ansatz is not a suitable model for the problem, considering its maximal mean correct rate of $77.30\%$ with full label information.

With this toy model, we demonstrate that QAL is valid if there is an adequate QNN with sufficient learnability, as well as the trade-off between expressivity and equivariance. Moreover, it should also work for binary classifying datasets with either discrete or continuous label symmetry if the distribution of classical data $\textbf{x}_i$ in domain $\mathcal{X}$ is continuous (e.g., donut-like datasets) or the outputs from quantum experiments $\rho_i$ are living in the whole Hilbert space (e.g., classifying single-qubit density matrices according to purity). In this way, the model can easily select the samples that are nearest to the decision boundary for labeling, making the optimized sampling behaviors more effective than sampling over those with almost the same distances to the decision boundary.

\section{tic-tac-toe: $\mathcal{D}_4$ symmetry}

\label{tic-tac-toe}

\begin{figure*}
	\includegraphics[width=0.95\linewidth]{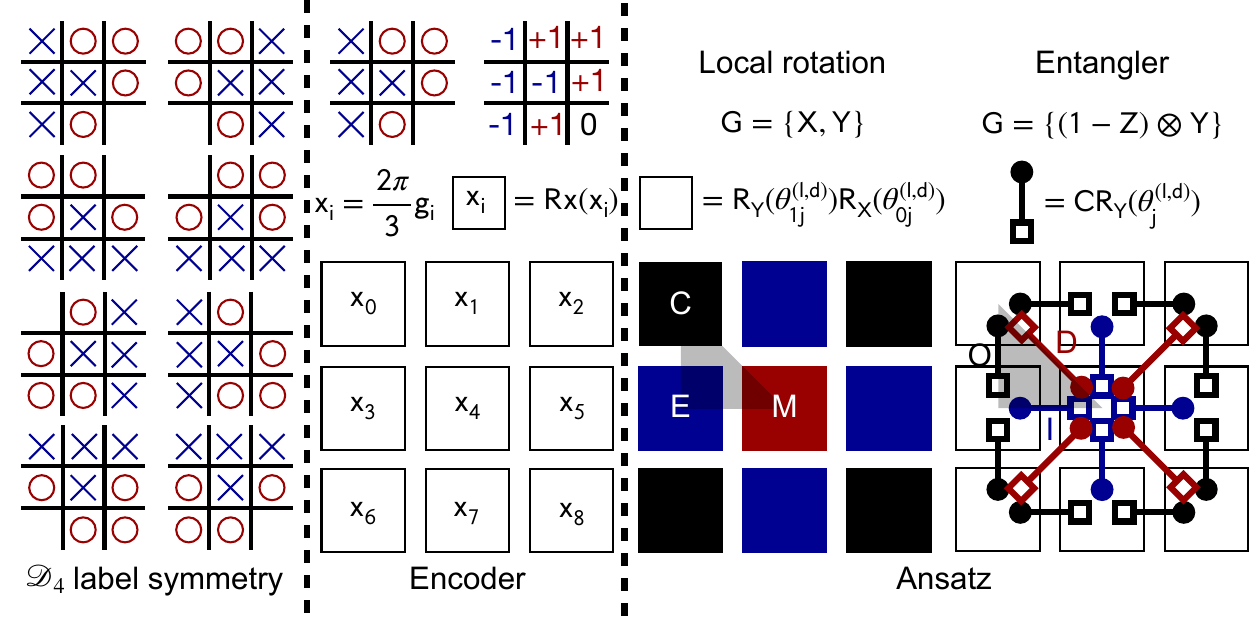}
	\caption{\label{fig:d4} In tic-tac-toe games, the winner is independent of rotations and flips, indicating a label symmetry of the dihedral group $\mathcal{D}_4$. Chess boards $\textbf{g}$ can be digitized into a sequence of integers among $g_i\in\{1,0,-1\}$ for QML tasks. The encoder (denoted by "T") consists of $R_x$ gates with parameters $x_i=\frac{2}{3}g_i$ for data encoding. The ansatz with geometric priors is comprised of single-qubit gates ($R_X$ and $R_Y$) on the corner (black "C"), edge (blue "E"), and middle (red "M"), where gates in the same category, repetition, and layer share a trainable parameter. Controlled $R_Y$ gates act as entanglers that connect corners to edges (black "O"), edges to middle (blue "I"), and middle to corner (red "D"), constructing the shadowed triangle to cover the qubit grid with label symmetry. The QNN $(l,d)$ has $l$ layers of trainable blocks, each of which starts with an encoder "T", followed by $d$ permutations of the ansatz "CEMOID".}
\end{figure*}

\subsection{Problem formulation}
After exploring QAL with continuous data encoding, we  study its performance with discrete data encoding using \textit{tic-tac-toe} as another toy model. The tic-tac-toe board has three classes: "cross $\times$ won," "circle $\circ$ won," and "draw $-$", determined by checking if any player ($\times$ or $\circ$) has formed a sequence of the same mark by row, column, or diagonal. Once again, Alice has results from 500 tic-tac-toe games,  divided into a ratio of $3:1:1$ for the training pool, validation set, and test set. Alice requires Bob to train a QML model for this multinomial classification problem with limited labeled samples selected from the pool. Ironically, Bob lacks knowledge of tic-tac-toe,  preventing him from inferring the winner despite having full access to the configurations of all chess boards. Thus, similar to the approach used in \textit{slicing-a-donut}, he relies on Alice to provide label information whenever  a chess board is chosen from the pool.

\subsection{QML model}
Although Bob lacks common sense in classical deterministic method for classification, he is an expert in group theory. He decides to incorporate the symmetry of the game into the model design. As shown on the left side of Fig.~\ref{fig:d4}, the label of an arbitrary chessboard is preserved under rotations of ${0, \pi/2, \pi, 3\pi/2}$ and the corresponding flipping about the vertical axis. Bob understands that this symmetry is characterized by the dihedral group $\mathcal{D}_4$, with an order of 8.

This symmetry ensures that corners remain mapped to corners, edges to edges, and the middle remains unchanged after the operation. It inspires the use of one-hot encoding, which maps each site on the chessboard to one qubit, simplifying the encoding-induced representation of label symmetry $\mathcal{D}_4$ to just $\text{SWAP}$. The chessboard is digitized as $\textbf{g}=\{g_0,...,g_8\}$, where the status of each site is given as $g_i=\{+1,-1,0\}$, corresponding to ${\times,\circ,-}$, respectively. These are then mapped to rotation angles $x_i=2\pi g_i/3$ for the $R_X(x_i)$ gate, see the middle of Fig.~\ref{fig:d4}. The encoder layer, denoted as "T" in the QNN structure, appears repeatedly in the circuit for data re-uploading. Consequently, the permutation-type symmetry leads to equivariant gates with visualizable symmetry on the grid of 9 qubits. For example, a rotation of $\pi/2$ angles should preserve labels after the following mappings: $0\rightarrow6\rightarrow8\rightarrow2$ for corners and $1\rightarrow3\rightarrow7\rightarrow5$ for edges. This corresponds to the $\text{SWAP}_{06}\text{SWAP}_{28}\text{SWAP}_{08}$ and $\text{SWAP}_{37}\text{SWAP}_{15}\text{SWAP}_{35}$ operations, respectively. Similarly, a vertical flip $0\leftrightarrow2$, $3\leftrightarrow5$, $6\leftrightarrow8$ corresponds to $\text{SWAP}_{02}\text{SWAP}_{35}\text{SWAP}_{68}$. Therefore, single-qubit gates should act on qubits of the same category with the same parameters, i.e., $R_Y(\theta_{1j}^{(l,d)})R_X(\theta_{0j}^{(l,d)})$, where $j=\{"\text{C}","\text{E}","\text{M}"\}$ represents the qubit on the corner (C), edge (E), and middle (M). The entangling gate should also follow this symmetry, e.g., $CR_Y(\theta_j^{(l,d)})$, linking all neighboring corners to edges symmetrically, leading to corner-to-edge (O), edge-to-middle (I), and middle-to-corner (D). The gate symmetry is illustrated on the right side of Fig.~\ref{fig:d4}. The correctness of the gateset can be verified by calculating the twirling with reduced symmetry, ensuring that symmetry between qubit 0 and qubit 2, for instance, is satisfied. For the single-qubit gate $R_X$ with generator $X_0$, we have the twirling formula
\begin{eqnarray}
T_R[X_0] &=& \frac{1}{2}[IX_0I+\text{SWAP}_{02}X_0\text{SWAP}_{02}]\nonumber\\
&=& (X_0+X_2)/2,
\end{eqnarray}
giving $R_X$ gate on the qubit 0 and qubit 2 by $\exp(-i\theta T_R[X_0])$, absorbing the all coefficients in gate parameters. Meanwhile, for the two-qubit gate $CR_Y$ with qubit 4 in the middle as the control qubit and the generator $G=(1-Z)_4\otimes Y_0$, the formula reads
\begin{eqnarray}
T_R[G] &=& \frac{1}{2}\{I_{402}[{(I-Z)_4\cdot\otimes Y_0\otimes I_2}] I_{402}\nonumber\\
&+&{[I_4\otimes\text{SWAP}_{02}]}[{(I-Z)_4\otimes Y_0\otimes I_2}]{[I_4\otimes\text{SWAP}_{02}]}\}\nonumber\\
&=& \frac{1}{2}[(I-Z)_4\otimes Y_0\otimes I_2+(I-Z)_4\otimes I_0 \otimes Y_2],
\end{eqnarray}
generating $CR_Y$ gates that target qubit 0 and 2 with the same parameters. The QNN with geometric prior consists of $l$ layers, each of which has one encoder "T" followed by $d$ repetitions of the equivariant ansatz "CEMOID". Note that the ansatz configuration can be permutations of "CEMOID" without loss of equivariance. In numerical experiments, we take $(l,d)=(2,5)$ with the QNN configuration denoted by "TCEMOIDCEMOIDCEMOIDCEMOIDCEMOIDTCEMOIDCEMOIDCEMOIDCEMOIDCEMOID". After applying the QNN to $|0\rangle^{\otimes 9}$, the prediction for each class is obtained by measuring equivariant observables
\begin{eqnarray}
\hat{\mathcal{O}}_{\circ} &=& \frac{1}{4}\sum_{\text{C}}Z_j=\frac{1}{4}[Z_0+Z_2+Z_6+Z_8],\\
\hat{\mathcal{O}}_{-} &=& Z_{\text{M}} = Z_4,\\
\hat{\mathcal{O}}_{\times} &=& \frac{1}{4}\sum_{\text{E}}Z_j=\frac{1}{4}[Z_1+Z_3+Z_5+Z_7]
\end{eqnarray}
for the vector of expectation values $\hat{\textbf{y}}=[\langle\hat{\mathcal{O}}_{\circ}\rangle,\langle\hat{\mathcal{O}}_{-}\rangle,\langle\hat{\mathcal{O}}_{\times}\rangle]$, where each expectation value is within $[-1,1]$, and finding the maximal expectation.

\subsection{Numerical experiments}

Before applying QAL, we need to evaluate the model's performance with all samples labeled from the training pool. We use the mean square error (MSE) loss function:
\begin{equation}
\label{eq:loss_mse}
\mathcal{L}(\tilde{\mathcal{U}},\boldsymbol{\theta}) = \frac{1}{\tilde{\mathcal{U}}}\sum_{\textbf{g}_i\in\tilde{\mathcal{U}}} ||\hat{\textbf{y}}(\textbf{g}_i,\boldsymbol{\theta})-\textbf{y}_i||^2,
\end{equation}
where $\textbf{y}_i$ is the one-hot encoded label of the chess board $\textbf{g}_i$ that $+1$ is assigned to the correct class and $-1$ to the others. The loss function is minimized using the Adam optimizer with a learning rate of 0.1 for 200 epochs without batching. All settings, including dataset separation, parameter initialization method, and stopping criterion, remain consistent with those used in the \textit{slicing-a-donut} experiment. We opt for the Mean Squared Error (MSE) loss function instead of cross-entropy to expedite the learning process with fewer epochs and a higher learning rate. We have trained the model with 40 different random parameter initializations over all labeled chess boards in the training pool. Achieving an accuracy of $76.89\pm6.54\%$ after training for an average of $127.58\pm61.20$ epochs, we conclude that the hyperparameter settings are adequate and the model is capable of classifying tic-tac-toe chess boards. It also shows a good agreement with the results from other numerical experiments, where this family of equivariant QNN was firstly proposed~\cite{meyer2023exploiting}. It is noteworthy that one can cherry-pick an accuracy of $88.89\%$ among the 40 results, significantly outperforming the $33.33\%$ baseline, which corresponds to blind guessing in a triple classification problem.

\begin{figure}
\includegraphics[width=0.95\linewidth]{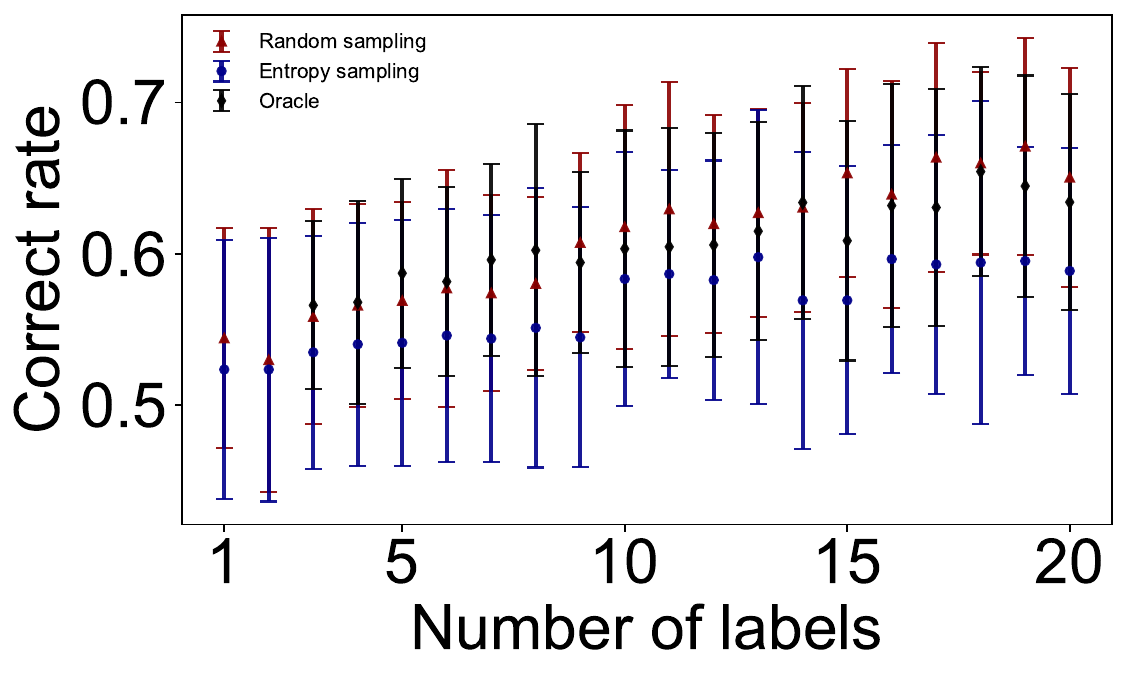}
\caption{\label{fig:entropy} Mean correct rate of quantum machine learning models with random sampling, entropy sampling, and entropy sampling with an initial training set of unbiased oracles as different sampling strategies for classifying \textit{tic-tac-toe} games. The error bar denote one standard deviation.}
\end{figure}

For triple classification, these sampling strategies are no longer equivalent. Additionally, it is necessary to map the vector of expectation values to a probabilistic distribution for USAMP. This can be achieved by employing the softmax function
\begin{equation}
P_{y_i} = \frac{\exp(y_i)}{\exp(\langle\hat{\mathcal{O}}_{\circ}\rangle)+\exp(\langle\hat{\mathcal{O}}_{-}\rangle)+\exp(\langle\hat{\mathcal{O}}_{\times}\rangle)},
\end{equation}
where $y_i$ are the expectation values in the vector. Thus, one can take the probabilities of all classes into consideration when assessing uncertainty, sampling by entropy
\begin{equation}
\textbf{g}_{\text{ES}} = \underset{\textbf{g}}{\text{argmax}}\left[-\sum_jP_\theta(y_j|\textbf{g})\log P_\theta(y_j|\textbf{g})\right]
\end{equation}
to query the more informative chess board $\textbf{g}$ according to the model estimation. In Fig.~\ref{fig:entropy}, we demonstrate QAL for tic-tac-toe with margin sampling over 40 different parameter initializations. Benchmarking against random sampling, QAL fails to outperform the baseline after querying 20 chess boards in the training pool.

The ineffectiveness of QAL in the tic-tac-toe scenario can be attributed to several factors. Firstly, tic-tac-toe chess boards may lack significant variation in terms of their informativeness. Even if a specific board is deemed more informative by the model, the reduction in uncertainty upon labeling it may not be substantial compared to labeling other boards. Additionally, in multinomial classification tasks with high-dimensional input spaces like tic-tac-toe, observations associated with a particular class may be distributed across disconnected decision subspaces. This complexity leads to biased sampling behavior, akin to focusing on the surface of a single bubble without considering the broader distribution within the input space. In our experiments, we observe that QAL primarily queries samples belonging to classes "cross" and "circle," neglecting "draw" samples. This bias leads to an unbalanced training set, ultimately resulting in reduced performance compared to random sampling, which ensures a more diverse selection of samples across classes.

\begin{figure}[t]
\includegraphics[width=1\linewidth]{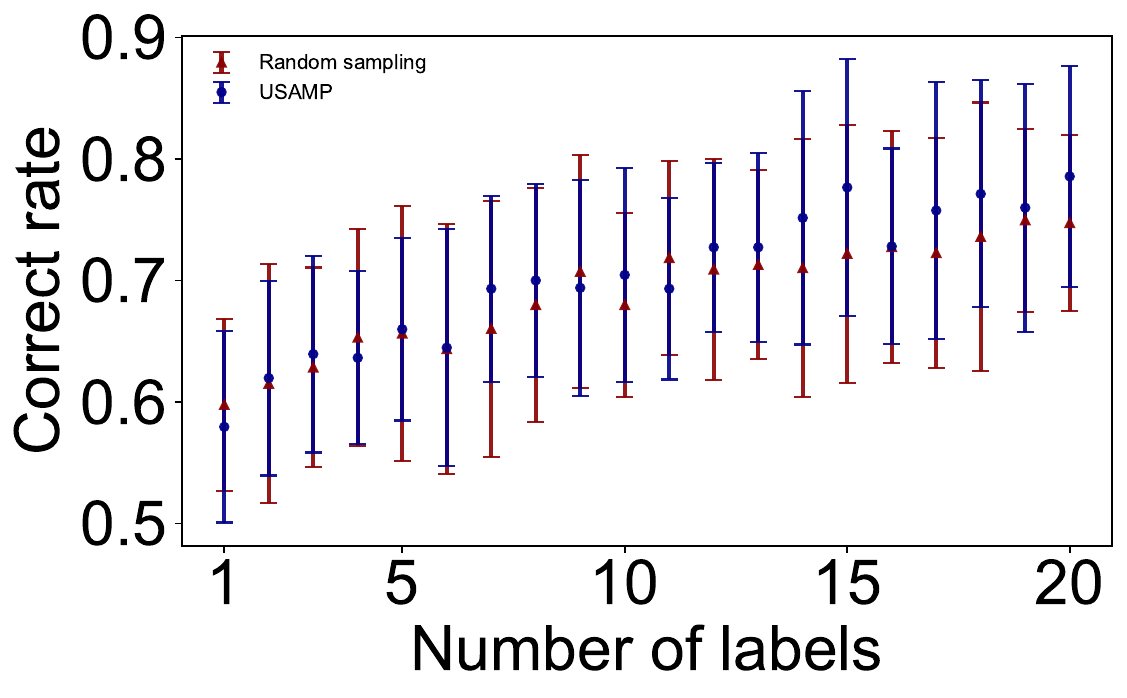}
\caption{\label{fig:binary} Mean correct rate of quantum machine learning models with random sampling and uncertainty sampling for classifying \textit{tic-tac-toe} games with winners as binary classification. The error bar denote one standard deviation.}
\end{figure}

To further investigate our hypothesis, Bob initiates with three chess boards representing each class as an oracle to initialize the model. Later, Bob employs entropy sampling and compares its performance with random sampling (depicted by black diamonds in Fig~\ref{fig:entropy}). Interestingly, entropy sampling with oracles initially outperforms random sampling, highlighting the benefits of including informative samples. However, as more samples are added to the training set, random sampling eventually surpasses QAL due to its biased sampling behavior. In Fig.~\ref{fig:binary}, we consider a reduced version of the problem focusing only on classifying chess boards with certain winners ($\times$ or $\circ$). Although the loss function remains unchanged, we modify the observation vector $\hat{\textbf{y}}$ to only include the expectation values of these two classes. In this unbiased scenario, QAL demonstrates clear advantages over the baseline, confirming our hypothesis regarding the failure of QAL in multinomial classification tasks. These additional experiments provide further support for our argument.

\section{Conclusion and outlook}

\label{Conclusion}

In summary, we have introduced QAL as an approach to achieve efficient QML with a limited number of samples. QNNs are employed to estimate the uncertainty of unlabeled samples in the pool, subsequently expanding the training set through iterative labeling of queried samples by human annotators. By integrating this formalism with GQML, the QNNs with geometric priors introduce a trade-off between equivariance and expressibility, leading to significant enhancements in model performance with a minimal number of samples. Our methodology has been exemplified through numerical experiments conducted in two distinct scenarios: one involving \textit{slicing-a-donut} characterized by $\mathcal{Z}_2$ symmetry for binary classification, and the other focusing on \textit{tic-tac-toe} with $\mathcal{D}_4$ symmetry for triple classification. While benchmarked against random sampling, our approach yields both positive and negative outcomes, underscoring the nuanced nature of QAL. We have delved into an analysis of the negative results, which is corroborated by additional specifically designed numerical experiments. Our findings indicate that the efficacy of QAL hinges on the appropriateness of the querying behavior. It is apparent that standard sampling strategies are not the silver bullet for all problems, as they may engender biased sampling behaviors that compromise model performance. This phenomenon is not unique to QAL and has been observed in classical machine learning, particularly in Bayesian models applied to multinomial classification tasks with quantum data~\cite{ding2022active}. To mitigate such biases, it is imperative to optimize sampling strategies, such as employing density-weighted sampling for datasets with non-standard distributions. Therefore, an intriguing avenue for future research entails exploring QAL strategies that mitigate the biased sampling behaviors illustrated here, which goes beyond the scope of this study.

A natural extension of this study lies in the application of QAL to quantum data, particularly data generated from quantum experiments. QAL can effectively collaborate with GQML, as quantum data commonly possesses label symmetries dependent on the classification method employed. Therefore, leveraging this synergy can notably enhance the performance of QNNs equipped with geometric priors, specifically tailored for analyzing outputs from both conventional and quantum-enhanced experiments.

An additional avenue worth exploring involves integrating QAL with various other QML models. While GQML offers an effective approach for enhancing few-shot learning capabilities, it may not always capture all relevant symmetries present in unlabeled datasets. Thus, it would be intriguing to investigate the potential of alternative QML models, such as quantum support vector machines~\cite{rebentrost2014quantum}, quantum convolutional neural network~\cite{cong2019quantum}, or recurrent quantum neural networks~\cite{bausch2020recurrent}, for addressing diverse problems with limited samples. Furthermore, for applications requiring few-shot learning, meta-learning techniques~\cite{vilalta2002perspective} could prove valuable by optimizing classical parameters to initialize quantum neural networks efficiently. The feasibility of this proposal has been demonstrated in preliminary research regarding the quantum approximate optimization algorithm~\cite{chandarana2023meta}.

\textit{Code Availability.---}
We utilized the quantum platform MindSpore Quantum~\cite{mq_2021} in our experiments. All codes, datasets, and results are provided in this \href{https://gitee.com/mindspore/mindquantum/tree/research/paper_with_code/quantum_active_learning_with_geometric_insight}{repo} for reproducibility of this study. Credit for the tic-tac-toe dataset generation code is attributed to PennyLane on their \href{https://pennylane.ai/qml/demos/tutorial_geometric_qml/}{webpage}.

\begin{acknowledgements}
The discussion on reproducing the preliminary results with Johannes Jakob Meyer, Francesco Arzani, and Cenk T\"uys\"uz is appreciated. This work is supported by NSFC (12075145 and 12211540002), STCSM (2019SHZDZX01-ZX04), the Innovation Program for Quantum Science and Technology (2021ZD0302302),  EU FET Open Grant EPIQUS (899368), QMiCS (Grant No. 820505), HORIZON-CL4-2022-QUANTUM-01-SGA project 101113946 OpenSuperQPlus100 of the EU Flagship on Quantum Technologies, the Basque Government through Grant No. IT1470-22, the Basque Government QUANTEK project under the ELKARTEK program (KK-2021/00070),  the project grant PID2021-126273NB-I00 and  PID2021-125823NA-I00 funded by MCIN/AEI/10.13039/501100011033, by ``ERDFA way of making Europe",  ``ERDF Invest in your Future", Nanoscale NMR and complex systems (PID2021-126694NB-C21, PID2021-126694NA-C22),  the Valencian Government Grant with Reference Number CIAICO/2021/184, the Spanish Ministry of Economic Affairs and Digital Transformation through the QUANTUM ENIA project call -- Quantum Spain project, and the European Union through the Recovery, Transformation and Resilience Plan--NextGenerationEU within the framework of the Digital Spain 2026 Agenda.   The authors also acknowledge the financial support received from the IKUR
Strategy under the collaboration agreement between Ikerbasque Foundation and BCAM on behalf of the Department of Education of the Basque
Government.  Y. D. acknowledges "CPS-MindSpore funding for quantum machine learning" with the support from the technical team of MindSpore Quantum.   X.C. and M. S. acknowledge ``Ayudas para contratos Ram\'on y Cajal'' 2015-2020 (RYC-2017-22482 and RYC-2020-030503-I).

\end{acknowledgements}

\end{document}